\documentclass[preprint,12pt]{elsarticle}

\usepackage{bm}
\usepackage{subfigure}
\usepackage{caption}
\usepackage{amssymb}

\journal{arxiv.org}

\begin{document}

\begin{frontmatter}



\title{Fractional constitutive equation (FACE) for non-Newtonian fluid flow: Theoretical description}
\author{HongGuang Sun}
\address{State Key Laboratory of Hydrology-Water Resources and Hydraulic Engineering, College of Mechanics and Materials, Hohai University, Nanjing 210098, China}
\author{Yong Zhang}
\address{1. College of Mechanics and Materials, Hohai University, Nanjing 210098, China\\
2. Department of Geological Sciences, University of Alabama, Tuscaloosa, AL 35487, United States\\
Corresponding author: yzhang264@ua.edu}
\author{Song Wei}
\address{College of Mechanics and Materials, Hohai University, Nanjing 210098, China}
\author{Jianting Zhu}
\address{Department of Civil and Architectural Engineering, University of Wyoming, 1000 E. University Ave., Laramie, WY 82071, United States}
\begin{abstract}
Non-Newtonian fluid flow might be driven by spatially nonlocal velocity, the dynamics of which can be described by promising fractional derivative models.  This short communication proposes a space FrActional-order Constitutive Equation (FACE) that links viscous shear stress with the velocity gradient, and then interprets physical properties of non-Newtonian fluids for steady pipe flow.  Results show that the generalized FACE model contains previous non-Newtonian fluid flow models as end-members by simply adjusting the order of the fractional index, and a preliminary test shows that the FACE model conveniently captures the observed growth of shear stress for various velocity gradients.  Further analysis of the velocity profile, frictional head loss, and Reynolds number using the FACE model also leads to analytical tools and criterion that can significantly extend standard models in quantifying the complex dynamics of non-Newtonian fluid flow with a wide range of spatially nonlocal velocities.
\end{abstract}
\end{frontmatter}
\section{Introduction}

Non-Newtonian fluids have many real-world applications \cite{Berli2008,Mahmood2009}.  Theoretical investigation provides useful information for the analysis and simulation of mass or energy transfer in non-Newtonian fluids \cite{Luikov1969,Li2011}, although their anomalous flow behaviors are usually more complex than Newtonian fluids.  Extensive experiments and measurements have revealed a nonlinear relationship between viscous shear stress and velocity gradient for non-Newtonian fluids such as muddy clay, oils, blood, paints, and polymeric solutions \cite{Tapadia2006,Pimenta2012}. Several empirical or semi-empirical formulas, such as the well-known power-law model, the Bingham model, and the Casson model, have been proposed to successfully quantify non-Newtonian viscosity behaviors observed in multiple disciplines \cite{Matsuhisa1965}.  The present models however suffer from several major drawbacks.  For example, they lack a unified constitutive description for most non-Newtonian fluids.  In addition, model parameters obtained from one flow system usually cannot be extended to another system for the same fluid \cite{Huilgol2016}.  This communication addresses the first challenge, with the expectation that a generalized constitutive equation may lead to transformative parameters.

Complex non-Newtonian fluids may be related to behaviors of their components, whose dynamics may exhibit complex memory impact in time or space.  Observations and measurements (for systems at either micro- or macro-scale) show that these fluids are often mixtures of materials with different sizes, such as water, solid particles, polymer, oil, and other long chain molecules \cite{Evans2015}.  The kinematic and dynamic behavior of these mixed materials can frequently exhibit long-time memory and (spatially) non-local properties \cite{Gibson1999,Siginer2014}.  The Newtonian constitutive equation only involves local influence of the system, which is captured by an integer-order velocity gradient in the model.  A generalization of the standard, integer-order constitutive equation might be necessary to incorporate memory and non-local properties for non-Newtonian fluids. Existing models including the power-law model and the shear rate dependent dynamic viscosity coefficient models can only partially describe the time history effect and non-local properties \cite{Berli2008,Karimi2014,Johnston2004,Johnston2006}. Moreover, shear rate dependent dynamic viscosity coefficient models usually generate nonlinear equations which are not easy to get the analytical expressions for real-world applications. Besides, shear rate dependent viscosity models have various expression forms for different kinds of non-Newtonian fluids, which have caused confusion in applications.

Fractional-derivative models were recently suggested by several studies as a promising tool to characterize non-Newtonian fluids, because fractional calculus can well characterize a physical system with long-term memory and spatial non-locality \cite{Ochoa-Tapia2007,Yin2012,SHAN2009,Ionescu2017}.  For example, the time fractional-derivative based constitutive equation, which can well describe the history dependency in fluid dynamics, was proposed to characterize different types of time-dependent non-Newtonian fluids \cite{Yin2012}.  This equation was further applied to analyze the various dynamic processes of muddy clay \cite{Yin2012}, seepage flow in dual-porosity media \cite{SHAN2009}, blood viscosity \cite{Ionescu2017}, and the behavior of Sesbania gel and xanthan gum \cite{Song1998}.  Previous studies of fractional-derivative models for non-Newtonian fluids, however, are mainly from the rheology viewpoint.  Most importantly, the possible spatial non-local dependency of flow velocity, which can significantly affect non-Newtonian fluid dynamics, has not been sufficiently addressed, although a few researchers have proposed that space fractional-derivative models may have the ability to capture non-Newtonian fluids.  For example, Ochoa-Tapia et al. \cite{Ochoa-Tapia2007} provided a brief mathematical derivation of fractional Newton's law for viscosity based on Taylor series, to obtain a fractional-order Darcy's law for describing shear stress phenomena in non-homogeneous porous media.  Chen et al. \cite{Chen2015} investigated the effects of model parameters to simulate the boundary layer flow of Maxwell fluids on an unsteady stretching surface, using the time-space fractional Navier-Stokes equation built upon a fractional-derivative based constitutive equation.  However, the physical analysis and description of non-Newtonian fluid flow based on the fractional-derivative constitutive equation, which addresses the non-locality of the velocity field (represented by a strong correlation between velocities), has not been investigated, hindering practical application of the space-fractional models.

This paper proposes and investigates a fractional-derivative constitutive equation (FACE) by employing a fractional-derivative velocity gradient, in which the space fractional-derivative is designed to address the non-local effect of velocity (likely due to complex interactions between non-uniform components and/or the impact of complex systems) on non-Newtonian fluid flow.  Application of the space fractional-derivative is motivated by the studies reviewed above.  We then use the FACE model to reclassify non-Newtonian fluids, and test its applicability using literature data.  Related physical quantities including velocity profile, frictional head loss, and fractional Reynolds number, are also analyzed for details of non-Newtonian fluid in steady pipe flow, to provide a theoretical description of non-Newtonian fluids necessary for real-world applications \cite{Berli2008,Eskin2008}.

\section{Fractional-derivative constitutive equation for non-Newtonian fluid}
The classical, empirical Newton's constitutive relationship for shear stress in terms of the velocity gradient can be expressed as
\begin{eqnarray}
\displaystyle{\tau=\mu \frac{d u}{d y},} \label{eq1}
\end{eqnarray}
in which $\tau$ is the viscous shear stress, $\mu$ is the dynamic viscosity, $u$ is the flow velocity, $du/dy$ is the velocity gradient, and $y$ is the direction perpendicular to the fluid flow.  Many kinds of fluids (such as slurry, blood, and rubber) however have been found to be non-Newtonian fluids which do not follow the classical Newton's law of viscosity (\ref{eq1}).

Previous investigations have offered several general extensions of Eq. (\ref{eq1}) for various non-Newtonian fluids. One of the most commonly used formulas is the following power-law model (which is also called the Ostwaald-de Waele model)
\begin{eqnarray}
\displaystyle{\tau=\tau_0+\mu \left( \frac{d u}{d y} \right)^{n},} \label{eq2}
\end{eqnarray}
in which $\tau_0$ represents the yield stress, $n$ is a parameter which corresponds to different kinds of non-Newtonian fluids. Although this model has proven useful in describing various kinds of non-Newtonian fluids, its rationality and physical origin remain obscure.  Extensive studies have shown that particle motion within non-Newtonian fluid has memory, and that the memory rate is related to the physical property of the target non-Newtonian fluid \cite{Irgens2014,Yin2012}.  To accurately describe time-dependent flow or rheology behavior of non-Newtonian fluids (or viscoelastic material) such as creep, the following time-fractional constitutive equation has been proposed \cite{Yin2012}
\begin{eqnarray}
\displaystyle{\tau(t)=\tau_0+\theta \: \lambda_\beta \: \frac{d^{\beta-1} \dot{\varepsilon}}{d t^{\beta-1}}\;, \;\;0\leq\beta\leq 1,} \label{eq3}
\end{eqnarray}
where $d^{\beta-1}/dt^{\beta-1}$ is the time fractional integral used to describe the time dependency of non-Newtonian fluids, $\theta$ and $\lambda$ are material constants, and $\dot{\varepsilon}$ denotes strain rate. This equation has been widely used to construct component models for Maxwell viscoelastic materials \cite{Koeller1984}, Oldroyd-B fluid \cite{Tong2005}, and some unsteady flows \cite{Tan2003}. 

To address the non-locality of non-Newtonian flow and the potential correlation of particles or components with different sizes, here we consider a general constitutive relationship by employing the fractional velocity gradient:
\begin{eqnarray}
\displaystyle{\tau=\tau_0+\mu_\alpha \: \frac{d^\alpha u}{d y^\alpha}\;, \;\;0<\alpha<2,} \label{eq4}
\end{eqnarray}
\begin{eqnarray}
\displaystyle{\tau=\mu_\alpha \: \frac{d^\alpha u}{d y^\alpha}\;, \;\;0<\alpha<2,} \label{eq4}
\end{eqnarray}
where $\alpha$ is the order of the space fractional derivative, and $d^\alpha u/d y^\alpha$ represents the fractional velocity gradient.  Here the physical origin of non-locality characterized by the fractional velocity gradient is likely a result of the mixing of non-uniform particles, close interaction of molecules, the existence of a continuous network of interactions between the elements \cite{Ovarlez2009}, or the effect of biological and chemical properties on non-Newtonian fluids \cite{Johnston2004}.  The definition of the fractional derivative used in this study is expressed by
\begin{eqnarray}
\begin{array}{c}
\displaystyle{\frac{d^\alpha u(y)}{d y^\alpha} =\frac{1}{\Gamma(n-\alpha)}\int_0^y \frac{u^{(n)}(\tau)}{(y-\tau)^{\alpha-n+1}}d \tau,\,\,n-1<\alpha \leq n,}
\end{array}
\label{eq5}
\end{eqnarray}
where $\Gamma(\cdot)$ denotes the Gamma function, and $n$ is the smallest integer greater than the order $\alpha$.  In this study, we consider a fractional derivative model for non-Newtonian fluid with $0<\alpha<2$.  In order to facilitate analysis, we further investigate the relationship between the velocity gradient and viscous shear stress.  Hence we rewrite Eq.(\ref{eq4}) as
\begin{eqnarray}
\tau=\left\{
\begin{array}{c}
\displaystyle{\tau_0+\mu I^{1-\alpha}(\frac{d u}{d y}),\,\,0<\alpha<1,} \\
\displaystyle{\tau_0+\mu \frac{d u}{d y},\,\,\alpha=1,} \\
\displaystyle{\tau_0+\mu I^{2-\alpha}[\frac{d }{d y}(\frac{d u}{d y})],\,\,1<\alpha<2,}
\end{array}\right.\;
\label{eq6}
\end{eqnarray}
where $I^{1-\alpha}$ and $I^{2-\alpha}$ represent the fractional integral
\begin{eqnarray}
\displaystyle{I^{\gamma}f(y)=\frac{1}{\it\Gamma(\gamma)} \int_{0}^y
(y-\tau)^{\gamma-1}f(\tau){\rm d}\tau,\quad \gamma>0.}
\label{eq6b}
\end{eqnarray}
A reclassification based on the FACE (\ref{eq4}) for non-Newtonian fluids is given in Fig. \ref{tab1} and Table \ref{tab1}.  Fig. \ref{tab1} clearly shows that the distinction of different types of fluids can be well characterized using the order of the fractional derivative.  It is also noteworthy that the limiting case, $\alpha=0$ represents elastomer and idea fluid when $\mu=0$. 

\begin{figure}[t]
\begin{center}
\setlength{\unitlength}{0.012500in}%
{\includegraphics[width=1.0\textwidth]{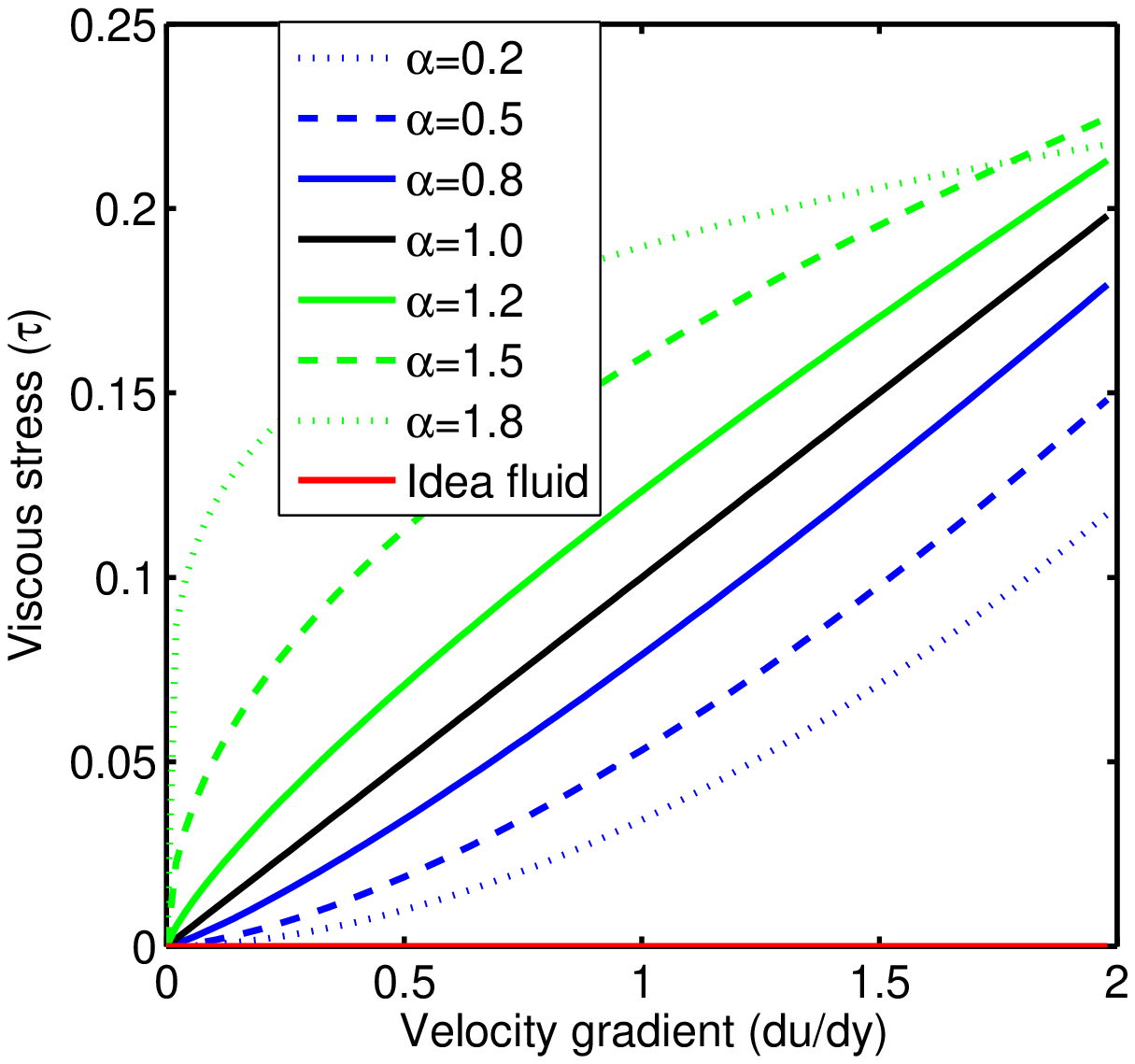}}\\
\end{center}
 \caption{The schematic diagram of relationship between viscous shear stress and velocity gradient with different yield stress and fractional derivative orders.  Other model parameters are: $du/dy=C y,\,\,C=2.0$, and $\mu=0.1$.
}
 \label{figure1}
\end{figure}

  \begin{table}
  \centering
   \caption{A reclassification of non-Newtonian fluids based on the FACE model (\ref{eq4}).}
\begin{tabular}{c c c}
   \hline
   yield stress ($\tau_0$) & Fractional derivative order ($\alpha$)
 &  Name of fluid \\
   \hline
  $\tau_0=0$ & $\alpha=0$ & Elastomer \\
  $\tau_0=0$ & $0<\alpha<1$ & Dilatant fluid \\
  $\tau_0=0$ & $\alpha=1$ &  Newtonian fluid \\
  $\tau_0=0$ & $2>\alpha>1$ & Pseudoplastic fluid \\
  $\tau_0=C (C>0)$ & $\alpha=1$ & Bingham Fluid (I) \\
  $\tau_0=C (C>0)$ & $0<\alpha<1$ & Bingham Fluid (II)\\
   \hline
 \end{tabular}
 \label{tab1}
  \end{table}

\begin{figure}[t]
\begin{center}
\setlength{\unitlength}{0.012500in}%
{\includegraphics[width=1.0\textwidth]{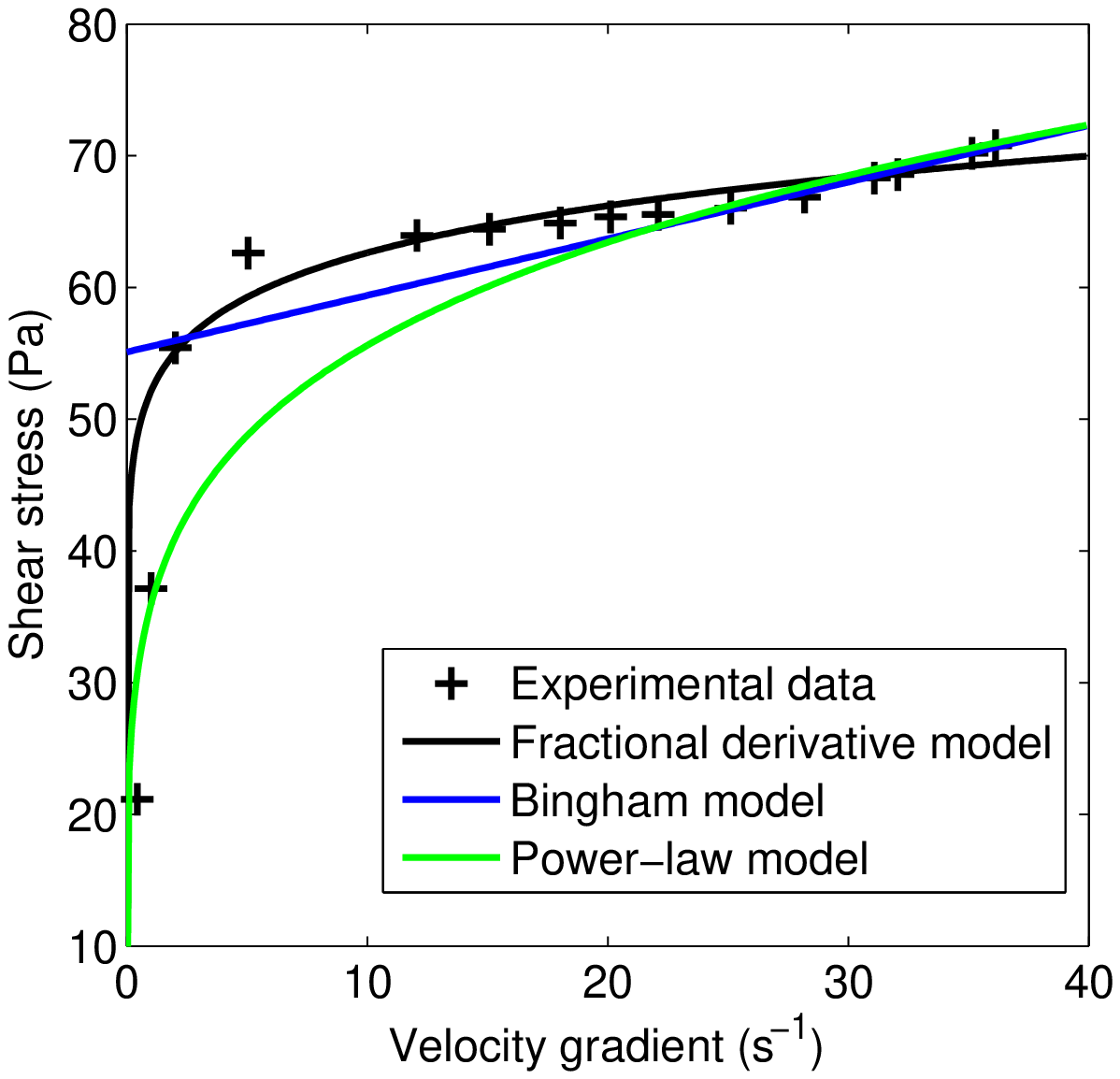}}\\
\end{center}
 \caption{Three models to fit the experimental data \cite{Salmon2003} of steady-state flow for a $6\%$ wt. CPCL/NaSal solution in $0.5$ M brine at $21.5^\circ C$.  The three models are: the fractional derivative model $\tau=50.0 d^\alpha u/dy^\alpha$, the power-law model $\tau=35.9 [u'(y)]^{0.19}$, and the Bingham model $\tau=55.1+0.43 u'(y)$.
}
 \label{figure2}
\end{figure}

For illustration purpose, we analyze a group of experimental data of wormlike micelles which have been found to be non-Newtonian fluid \cite{Salmon2003}, and further explore the relationship between shear stress and velocity gradient (or shear rate) using the FACE model (\ref{eq4}).  Fig. \ref{figure2} compares fitting results using the FACE model (\ref{eq4}), the power-law model, and the Bingham model.  It is clear that the power-law and Bingham models give good agreement at a high velocity gradient regime, while the FACE model (\ref{eq4}) provides accurate description for the entire region.

\section{Theoretical analysis of steady pipe flow}
For application purposes, we conduct theoretical derivations for major physical quantities related to non-Newtonian fluid in steady pipe flow.

\subsection{Velocity profile of pipe flow}
Here we assume that 1) the non-Newtonian fluid is incompressible, and 2) the pipe flow is laminar.
\begin{figure}[t]
\begin{center}
\setlength{\unitlength}{0.012500in}%
{\includegraphics[width=1.0\textwidth]{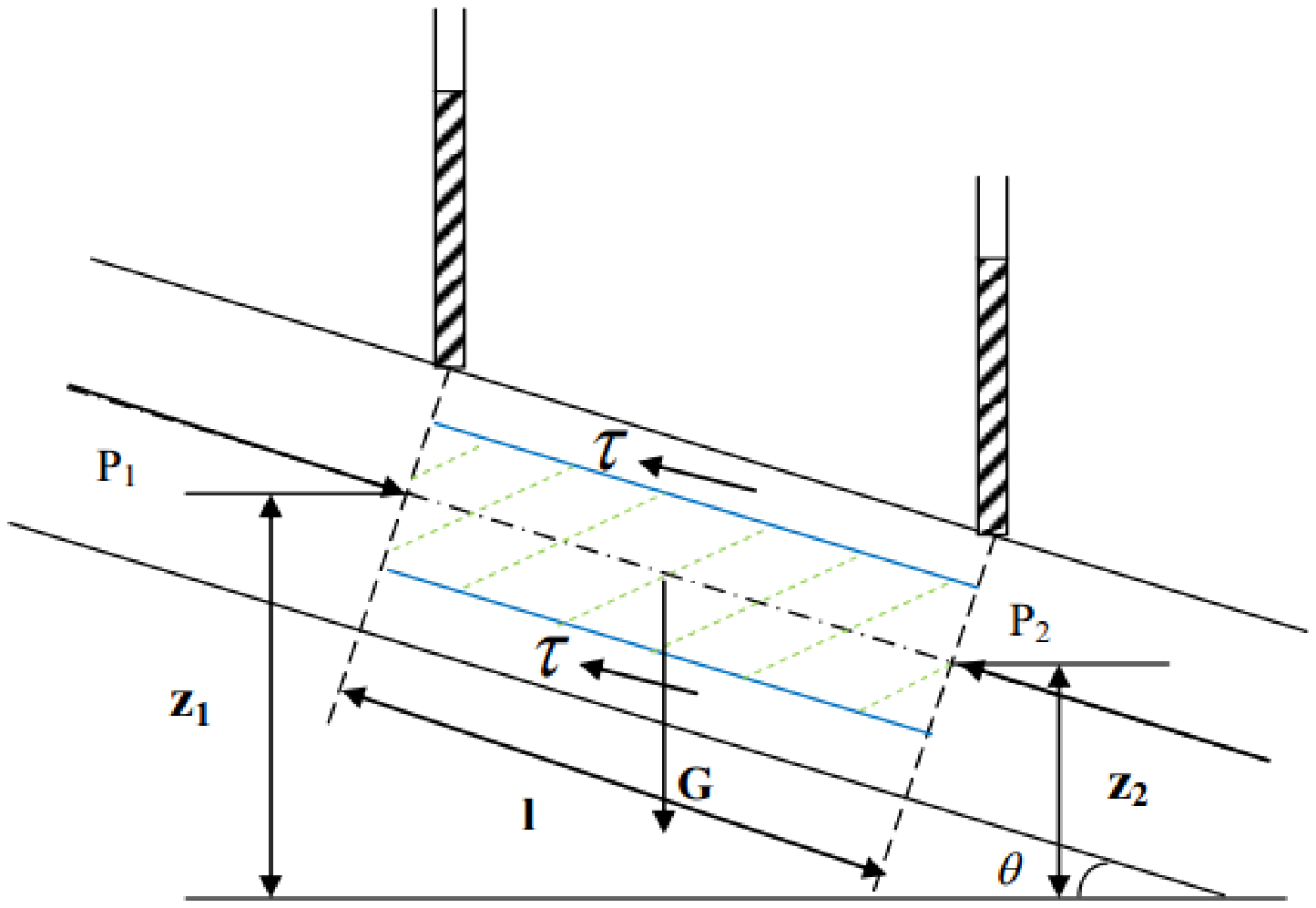}}\\
\end{center}
 \caption{Mechanical analysis of a fluid system in steady pipe flow.
}
 \label{figure3}
\end{figure}
Based on the mechanical analysis of a fluid system (shown by the cylinder with blue color in Fig. \ref{figure3}) in steady flow, the following force balance equation can be established in the flow direction
\begin{eqnarray}
\displaystyle{p_1 A'-p_2 A'+\rho g A' l sin \theta- \tau \cdot 2 \pi r l=0,} \label{eq7}
\end{eqnarray}
where $A'$ denotes the cross section area, $r$ is the radius of the cross section of the considered cylinder system, $\rho$ is the density of non-Newtonian fluid, and $R$ is the pipe radius.  Using the relationship of $sin \theta=(z_1-z_2)/l$, we can re-arrange (\ref{eq7}) to get
\begin{eqnarray}
\displaystyle{\frac{(z_1+p_1/\rho g)-(z_2+p_2/\rho g)}{l} = \frac{2 \tau}{\rho g r}.} \label{eq8}
\end{eqnarray}
Finally the viscous shear stress can be expressed as
\begin{eqnarray}
\displaystyle{\tau=\rho g r J/2,} \label{eq9}
\end{eqnarray}
where $J$ ($=h_f/l$) is the hydraulic gradient, and $h_f$ is the frictional head loss expressed by $h_f=(z_1+p_1/\rho g)-(z_2+p_2/\rho g)$.

It is clear that the viscous shear stress calculated by the fractional Newtonian constitutive equation Eq.(\ref{eq4}) should be equal to Eq. (\ref{eq9}) obtained by mechanical analysis. For illustration purposes, firstly we consider the non-Newtonian fluid without the yield stress $\tau_0$ (i.e., $\tau_0=0$), which leads to the following result
\begin{eqnarray}
\displaystyle{\mu \frac{d^\alpha u}{d r^\alpha}=-\rho g r J/2.} \label{eq10}
\end{eqnarray}
The negative sign on the right-hand side of (\ref{eq10}) comes from the opposite directions of two stress expressions. By employing the property of the Caputo fractional derivative \cite{Li2007}, we get
\begin{eqnarray}
\displaystyle{u(r)-u(r=0)=-\frac{\rho g J r^{1+\alpha}}{2 \mu \Gamma (\alpha+2)}.} \label{eq11}
\end{eqnarray}
Since flow velocity at the pipe wall is zero, we have the boundary condition $u(r=R)=0$, and then we get the following result
\begin{eqnarray}
\displaystyle{u(r=0)=\frac{\rho g J R^{1+\alpha}}{2 \mu \Gamma (\alpha+2)}.} \label{eq12}
\end{eqnarray}
Finally, the velocity profile is written as
\begin{eqnarray}
\displaystyle{u(r)=\frac{\rho g J }{2 \mu \Gamma (\alpha+2)}(R^{1+\alpha}-r^{1+\alpha}).} \label{eq13}
\end{eqnarray}
We emphasize that the above result is correct for both $0<\alpha\leq 1$ and $1<\alpha<2$, since $u_{max}=u(r=0)$ and $u'(r=0)=0$. It is also clear that Eq. (\ref{eq13}) reduces to the velocity profile of Newtonian fluid when $\alpha=1$.

\begin{figure}[htb]
\centering
\subfigure{
\includegraphics[width=0.5\linewidth]{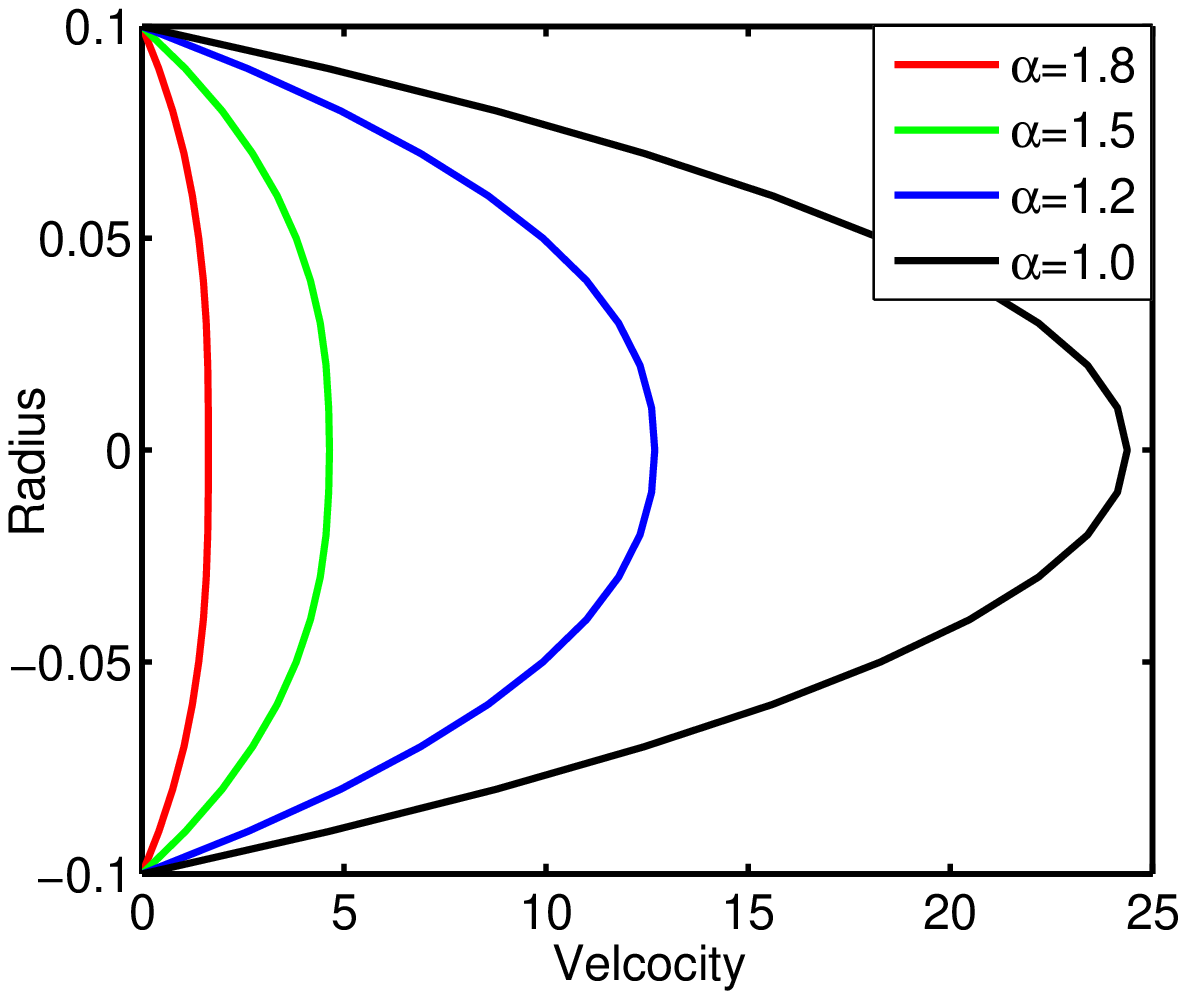}
\label{fig:forceBalanceComparison_a}}
\subfigure{
\includegraphics[width=0.5\linewidth]{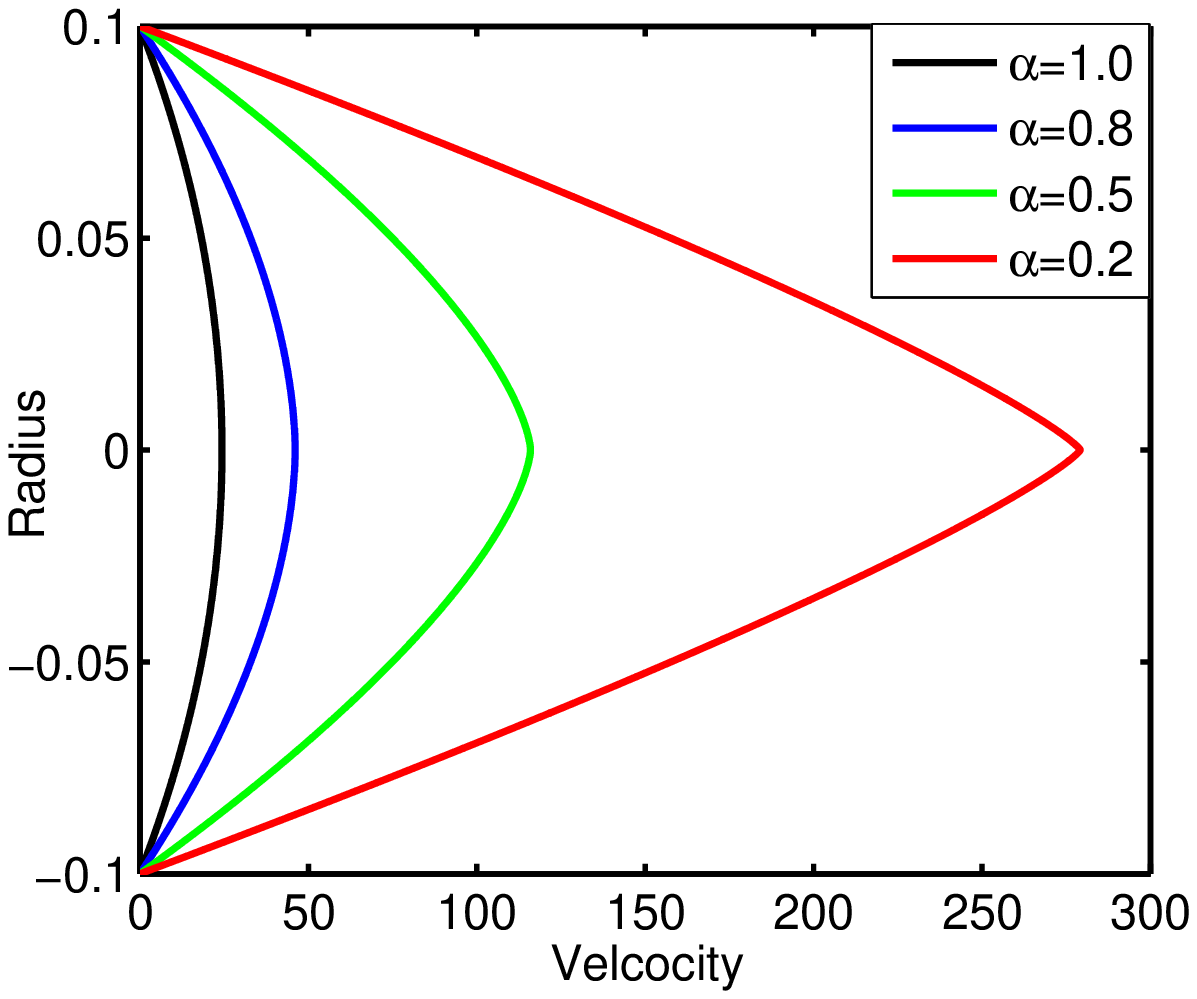}
\label{fig:forceBalanceComparison_b}}
\caption{Velocity profiles for non-Newtonian fluids in steady pipe flow, based on Eq. (\ref{eq13}).  The other model parameters (that remain constant in all cases) are: $g=9.8$ m/s$^2$ $\rho=1.0\times 10^3$ kg/m$^3$, $J=10^{-3}$, and $\mu=1.005\times 10^{-3}$ N$\cdot$ s/m$^{2-\alpha}$, which are the physical properties of water at $20^\circ C$.  Here the negative sign of radius is only used for illustration convenience.} \label{figure4}
\end{figure}

Moreover, the maximum velocity and mean velocity can be derived from Eq. (\ref{eq13})
\begin{eqnarray}
\left\{
\begin{array}{c}
\displaystyle{u_{max}=\frac{\rho g J R^{1+\alpha}}{2 \mu \Gamma (\alpha+2)},} \\
\displaystyle{\bar{u}=\frac{\rho g J }{2 \mu \Gamma (\alpha+2)}(1-\frac{2}{3+\alpha}) R^{1+\alpha},}
\end{array}\right.\;
\label{eq14}
\end{eqnarray}
and
\begin{eqnarray}
\displaystyle{\frac{\bar{u}}{u_{max}}=1-\frac{2}{3+\alpha},} \label{eq15}
\end{eqnarray}
in which $u_{max}$ and $\bar{u}$ represent the maximum and mean velocities in the cross section, respectively.  It is obvious that $\bar{u}/u_{max}<1/2$ when $0<\alpha<1$ and $\bar{u}/u_{max}>1/2$ if $1<\alpha<2$, and this ratio becomes $1/2$ when $\alpha=1$ (i.e., Newtonian fluid).  It implies that the velocity distribution of non-Newtonian fluid may be less or more uniform than Newtonian fluid, which agrees well with experimental data \cite{Vaughn1966}. A comparison of the velocity distribution of non-Newtonian fluids in steady pipe flow, as characterized by different fractional-derivative orders, is shown in Fig. \ref{figure4}.  Large differences in velocity distributions can be found when different fractional derivative orders are used.  In Fig. \ref{figure4}, we use the same physical parameters for different non-Newtonian fluids (for direct comparison).  It is noteworthy that those parameters only affect the velocity values, while the velocity distribution is determined mainly by the fractional order $\alpha$.  Further investigations of specific non-Newtonian fluids are needed in a subsequent study. 

\subsection{Frictional head loss}
Furthermore, we can get the expression of the frictional head loss ($h_f$) from the mean velocity in Eq. (\ref{eq14}):

\begin{eqnarray}
\displaystyle{h_f=\frac{2 \nu \Gamma(\alpha+2)}{g} \frac{3+\alpha}{1+\alpha} \frac{l \bar{u}}{R^{1+\alpha}},} \label{eq16}
\end{eqnarray}
in which $\nu=\mu/\rho$ is the kinematic viscosity.
Eq. (\ref{eq16}) can be rewrite as
\begin{eqnarray}
\displaystyle{h_f=\frac{64}{\frac{2^{3-\alpha}}{(3+\alpha) \Gamma(1+\alpha)}\frac{\bar{u} D^\alpha}{\nu}} \frac{l}{D} \frac{\bar{u}^2}{2g},} \label{eq17}
\end{eqnarray}
where $D=2 R$ is the diameter of the pipe.

\subsection{Fractional Reynolds number}
Eq. (\ref{eq17}) reduces to the expression of the frictional head loss for Newtonian fluid when $\alpha=1$:
\begin{eqnarray}
\displaystyle{h_f=\frac{64}{Re} \frac{l}{D} \frac{\bar{u}^2}{2g},} \label{eq18}
\end{eqnarray}
in which $Re=\bar{u} D/\nu$ is the Reynolds number of Newtonian fluid.  Combing (\ref{eq17}) and (\ref{eq18}), we hereby define the following Reynolds number for non-Newtonian fluid described by the FACE (\ref{eq4}):
\begin{eqnarray}
\displaystyle{Re_\alpha= \frac{2^{3-\alpha}}{(3+\alpha) \Gamma(1+\alpha)}\frac{\bar{u} D^\alpha}{\nu}.} \label{eq19}
\end{eqnarray}
This expression indicates that different types of non-Newtonian fluids own their critical Reynolds number, which is characterized by the fractional index $\alpha$ corresponding to specific non-Newtonian fluid.  For example, the critical Reynolds number of expansion fluid ($0<\alpha<1$) is larger than the pseudoplastic fluid ($1<\alpha<2$), when the diameter of the pipe is $D=1.0$ m.  Another interesting result is that the diameter $D$ plays a more important role in determining the fractional Reynolds number, in comparison with Newtonian fluid.  Extensive experiments are required in a future study to determine the feasibility of the generalized, fractional Reynolds number (\ref{eq19}) in distinguishing flow patters of various non-Newtonian fluids.

\subsection{Yield stress}
From a more general point of view, we further consider non-Newtonian fluid with yield stress $\tau_0$ (i.e., $\tau_0\neq 0$).  The velocity profile of pipe flow expressed by Eq. (\ref{eq13}) now should be changed to
\begin{eqnarray}
\displaystyle{u(r)=\frac{R^\alpha}{\mu \Gamma (1+\alpha)} \left[ \frac{\rho g J R}{2(1+\alpha)}-\tau_0 \right] -\frac{r^\alpha}{\mu \Gamma (1+\alpha)} \left[\frac{\rho g J r}{2(1+\alpha)}-\tau_0 \right].} \label{eq20}
\end{eqnarray}
This means that non-Newtonian fluid flow with nonzero-value yield stress occurs when the shear stress exceeds a critical value.  We also emphasize here that this expression (\ref{eq20}) is only valid for fully developed flow.  Various structural flows including plug flow may exist for variable viscous shear stress values, which can dramatically complicate the above analysis \cite{Ovarlez2009,Gabard2003}.  This remains an open research question.

\section{Conclusion}

This study proposes, explores, and tests the fractional constitutive equation (FACE) for describing non-Newtonian fluids.  Results show that the FACE model captures the most common behaviors of non-Newtonian fluid flow, and is sufficiently simple to allow for analytical expressions of the flow field in steady pipe flow.  Although lacking extensive experimental validation, the concept presented in this paper provides the first and fundamental step in an unpaved path to future development of reliable non-Newtonian fluid dynamic models using the promising fractional derivative.

\subsection*{Acknowledgment}
The work was supported by the National Natural Science Foundation of China (Grant Nos. 11572112, 41628202, and 11528205). This paper does not necessarily reflect the view of the funding agency. 

\subsection*{References}


\begin{thebibliography}{10}
\bibitem{Berli2008}
C.L.A. Berli, M.L. Olivares,  Electrokinetic flow of non-Newtonian fluids in microchannels, J. Colloid Interface Sci. 320(2)(2008) 582-589.

\bibitem{Mahmood2009}
A. Mahmood, S. Parveen, A. Ara, N.A. Khan, Exact analytic solutions for the unsteady flow of a non-Newtonian fluid between two cylinders with fractional derivative model,
Commun. Nonlinear Sci. Numer. Simul. 14(8)(2009) 3309-3319.

\bibitem{Luikov1969}
A.V. Luikov, Z.P. Shulman, B.I. Puris, External convective mass transfer in non-Newtonian fluid: Part I, Int. J. Heat \& Mass Tran. 12(4)(1969) 377-391.

\bibitem{Li2011}
B. Li, L. Zheng, X. Zhang, Heat transfer in pseudo-plastic non-Newtonian fluids with variable thermal conductivity, Energ. Convers. \& Manage. 52(1)(2011) 355-358.

\bibitem{Tapadia2006}
P. Tapadia, S.Q. Wang, Direct visualization of continuous simple shear in non-Newtonian polymeric fluids, Phys. Rev. Lett. 96(1)(2006) 016001.

\bibitem{Pimenta2012}
T.A. Pimenta, J. Campos,  Friction losses of Newtonian and non-Newtonian fluids flowing in laminar regime in a helical coil, Exp. Therm. \& Fluid Sci. 36(2012) 194-204.

\bibitem{Matsuhisa1965}
S. Matsuhisa, R.B. Bird, Analytical and numerical solutions for laminar flow of the non-Newtonian ellis fluid, AIChE J. 11(4)(1965) 588-595.

\bibitem{Huilgol2016}
R.R. Huilgol, G.H.R. Kefayati,  From mesoscopic models to continuum mechanics: Newtonian and non-newtonian fluids, J. Non-Newtonian Fluid Mech. 233(2016) 146-154.

\bibitem{Evans2015}
R.M.L. Evans, C.A. Hall, R.A. Simha, T.S. Welsh, Classical X Y model with conserved angular momentum is an archetypal non-Newtonian fluid, Phys. Rev. Lett. 114(13)(2015) 138301.

\bibitem{Gibson1999}
A.G. Gibson, S. Toll, Mechanics of the squeeze flow of planar fibre suspensions, J. Non-Newtonian Fluid Mech. 82(1)(1999) 1-24.

\bibitem{Siginer2014}
D.A. Siginer,  Stability of non-linear constitutive formulations for viscoelastic fluids, Springer, 2014.

\bibitem{Karimi2014}
S. Karimi, M. Dabagh, P. Vasava, M. Dadvar, B. Dabir, P. Jalali, Effect of rheological models on the hemodynamics within human aorta: CFD study on CT image-based geometry, J. Non-Newtonian Fluid Mech. 207(2014) 42-52.

\bibitem{Johnston2004}
B.M. Johnston, P.R. Johnston, S. Corney, D. Kilpatrick, Non-Newtonian blood flow in human right coronary arteries: steady state simulations, J. Biomech. 37(5)(2004) 709-720.

\bibitem{Johnston2006}
B.M. Johnston, P.R. Johnston, S. Corney, D. Kilpatrick, Non-Newtonian blood
flow in human right coronary arteries: transient simulations, J. Biomech. 39
(2006) 1116-1128.

\bibitem{Ochoa-Tapia2007}
J.A. Ochoa-Tapia, F.J. Valdes-Parada, J. Alvarez-Ramirez, A fractional-order Darcy's law, Phys. A 374(1)(2007) 1-14.

\bibitem{Irgens2014}
F. Irgens, Flow Phenomena \emph{in} Rheology and Non-Newtonian Fluids, Springer International Publishing (2014) 17-23.

\bibitem{Yin2012}
D. Yin, W. Zhang, C. Cheng, L. Yi,  Fractional time-dependent Bingham model for muddy clay, J. Non-Newtonian Fluid Mech. 187(2012) 32-35.

\bibitem{SHAN2009}
L. Shan, D. Tong, L. Xue,  Unsteady flow of non-Newtonian visco-elastic fluid in dual-porosity media with the fractional derivative, J. Hydrodyn. Ser. B 21(5)(2009) 705-713.

\bibitem{Ionescu2017}
C.M. Ionescu,  A memory-based model for blood viscosity, Commun. Nonlinear Sci. Numer. Simul. 45(2017) 29-34.

\bibitem{Song1998}
D.Y. Song, T.Q. Jiang,  Study on the constitutive equation with fractional derivative for the viscoelastic fluids-Modified Jeffreys model and its application, Rheol. Acta 37(5)(1998) 512-517.

\bibitem{Chen2015}
S. Chen, L. Zheng, B. Shen, X. Chen, Time-space dependent fractional boundary layer flow of Maxwell fluid over an unsteady stretching surface, Theor. Appl. Mech. Lett. 5(6)(2015) 262-266.

\bibitem{Eskin2008}
D. Eskin, M.J. Miller,  A model of non-Newtonian slurry flow in a fracture, Powder Tech. 182(2)(2008) 313-322.

\bibitem{Koeller1984}
R.C. Koeller,  Applications of fractional calculus to the theory of viscoelasticity, J. Appl. Mech. 51(2)(1984) 299-307.

\bibitem{Tong2005}
D. Tong, R. Wang, H. Yang, Exact solutions for the flow of non-Newtonian fluid with fractional derivative in an annular pipe, Sci. China Ser. G: Phys., Mech. Astron. 48(4)(2005) 485-495.

\bibitem{Tan2003}
W.C. Tan, W.X. Pan, M.Y. Xu,  A note on unsteady flows of a viscoelastic fluid with the fractional Maxwell model between two parallel plates, Int. J. Non-Linear Mech. 38(5)(2003) 645-650.

\bibitem{Ovarlez2009}
G. Ovarlez, S. Rodts, X. Chateau, P. Coussot, Phenomenology and physical origin of shear localization and shear banding in complex fluids, Rheol. Acta 48(8)(2009) 831-844.

\bibitem{Salmon2003}
J.B. Salmon, A. Colin, S. Manneville,  F. Molino, Velocity profiles in shear-banding wormlike micelles, Phys. Rev. Lett. 90(22)(2003) 228303.

\bibitem{Li2007}
C. Li, W. Deng,  Remarks on fractional derivatives, Appl. Math. \& Comput. 187(2)(2007) 777-784.

\bibitem{Vaughn1966}
R.D. Vaughn, P.D. Bergman, Laminar flow of non-Newtonian fluids in concentric annuli, Ind. Eng. Chem. Process Des. Dev. 5(1)(1966) 44-47.

\bibitem{Gabard2003}
C. Gabard, J.P. Hulin,  Miscible displacement of non-Newtonian fluids in a vertical tube, Eur. Phys. J. E 11(3)(2003) 231-241.

%

\end{thebibliography}
\end{document}